\documentclass[a4paper,11pt]{article}
\usepackage{pos}
\usepackage{subcaption}   
\usepackage{braket}
\usepackage{float}
\usepackage{bbold}
\usepackage{amsmath}

\newcommand{\ord}{\mathcal{O}}

\newcommand{\gfi}{\gamma_5}
\newcommand{\gam}{\gamma_\mu}

\newcommand{\gaumd}{g_A^{u-d}}

\title{Nucleon axial, tensor, and scalar charges and $\sigma$-terms in lattice QCD}

\author[a,b]{Constantia Alexandrou}
\author[b]{Simone Bacchio}
\author[c]{Jacob Finkenrath}
\author*[a]{Christos Iona}
\author[b]{Giannis Koutsou}
\author[a]{Yan Li}
\author[a]{Gregoris Spanoudes}

\affiliation[a]{Department of Physics, University of Cyprus, P.O. Box 20537, 1678 Nicosia, Cyprus}

\affiliation[b]{Computation-based Science and Technology Research Center, The Cyprus Institute, 20 Kavafi Str., Nicosia 2121, Cyprus}

\affiliation[c]{Department of Theoretical Physics, European Organization for Nuclear Research, CERN, CH-1211 Geneva 23, Switzerland}

\emailAdd{iona.n.christos@ucy.ac.cy}

\abstract{We determine the nucleon axial, scalar and tensor charges at the continuum limit  by analyzing three $N_f=2+1+1$
 twisted mass fermion ensembles with all quark masses tuned to approximately their physical values. We  include all contributions from valence and sea quarks. We use the Akaike Information Criterion to evaluate systematic errors due to excited states and the continuum extrapolation. For the nucleon isovector axial charge we find $g_A^{u-d}=1.250(24)$, in agreement with the experimental value. We compute the axial, tensor and scalar charges for each quark flavor. The axial charge provides  crucial information on  the intrinsic spin carried by quark in the nucleon and the the latter two provide input for experimental searches of physics beyond the standard model. Moreover, we extract the nucleon $\sigma$-terms and find  $\sigma_{\pi N}=41.9(8.1)$~MeV, for the strange $\sigma_{s}=30(17)$~MeV and for the charm $\sigma_{c}=82(29)$~MeV. We also present preliminary results on the isovector quantities using a fourth ensemble at smaller lattice spacing.}

\FullConference{The 41st International Symposium on Lattice Field Theory (LATTICE2024)\\
 28 July - 3 August 2024\\
Liverpool, UK\\}

\begin{document}
\maketitle

\section{Introduction}

The nucleon axial, tensor, and scalar charges, along with the $\sigma$-terms, are fundamental quantities that provide insights into nucleon structure. The axial charge, $\gaumd$, governs neutron beta decay and plays a critical role in neutrinoless double-beta decay and tests of CKM matrix unitarity. Flavor-diagonal axial charges, $g_A^f$, describe the intrinsic spin, $\frac{1}{2}\Delta \Sigma_q$, carried by quarks in the nucleon, as measured in deep inelastic scattering at facilities, such as Jefferson Lab and  CERN, and also targeted at the future Electron-Ion Collider (EIC).

The isovector tensor and scalar charges, $g_T^{u-d}$ and $g_S^{u-d}$, are crucial for constraining beyond the Standard Model (BSM) interactions to be searched in experiments like DUNE~\cite{Bischer:2018zcz}, COHERENT~\cite{COHERENT:2017ipa}, GEMMA~\cite{Beda:2012zz}, and dark matter detection searches~\cite{Undagoitia:2015gya,Chizhov:1995wp,Dhargyal:2016jgo}. Accurate determination of the tensor charge is key to understanding the transversity parton distribution function, while the nucleon $\sigma$-terms quantify quark mass contributions to the nucleon mass.

We compute the nucleon charges and $\sigma$-terms using twisted mass fermion ensembles at three lattice spacings and present preliminary results for the isovector charges at a fourth, finer lattice spacing. These $N_f=2+1+1$ ensembles, simulated by the Extended Twisted Mass Collaboration (ETMC), use quarks fixed near their physical masses~\cite{Alexandrou:2018egz}, allowing precise determinations without chiral extrapolations. 

\section{Nucleon Matrix Elements}

The nucleon axial, tensor, and scalar charges for each quark flavor $f$, denoted as $g^f_{\rm A,T,S}$, are extracted from the matrix elements of the corresponding axial, tensor, and scalar operators at zero momentum transfer:
\begin{equation}
\langle N|\bar{\psi}^f\Gamma_{\rm A,S,T}\psi^f|N\rangle = g^f_{\rm A,T,S} \bar{u}_N\Gamma_{\rm A,S,T} u _N\,,
\end{equation}
where $u_N$ is the nucleon spinor, and the operator structures are $\Gamma_A=\gamma_\mu\gamma_5$ (axial-vector), $\Gamma_S=\mathbb{1}$ (scalar), and $\Gamma_T=\sigma_{\mu\nu}$ (tensor), with $\sigma_{\mu\nu}=\frac{i}{2}[\gamma_\mu,\gamma_\nu]$. The renormalization group invariant $\sigma^f$-term is given by $\sigma^f=m_f\langle N|\bar{\psi}^f\psi^f|N\rangle$, with $m_f$ the mass of the quark with flavor $f$.

\begin{table}[ht!]
	\centering
	{\small
		\renewcommand{\arraystretch}{0.9}
		\renewcommand{\tabcolsep}{7.0pt}
        \caption{Parameters of the $N_f=2+1+1 $ ensembles analyzed in this work. In the first column, we give the name of
             the ensemble, in the second the abbreviated name, in the third the lattice volume, in the fourth $\beta = 6/g^2$ with $g$ the bare coupling constant, in the fifth
             the lattice spacing and in the sixth the pion mass. Lattice spacings and pion masses for B64, C80 and D96 are taken from Ref.~\cite{ExtendedTwistedMass:2022jpw} and for E112 from Ref.~\cite{ExtendedTwistedMassCollaborationETMC:2024xdf}. In the last column we list the number of configurations used per ensemble.}
        \label{tbl:Ensembles}
    \begin{tabular}{c|c|c|c|c|c|c}
    \hline \hline
        Ensemble     & Abrv. & $V/a^4$& $\beta$ & $a~[fm]$    & $m_\pi~[MeV]$ & Config. \\ \hline 
        cB211.072.64 & B64   & $64^3 \times 128$ & 1.778   & 0.07957(13) & 140.2(2)& 749     \\ 
        cC211.060.80 & C80   & $80^3 \times 160$ & 1.836   & 0.06821(13) & 136.7(2)& 400      \\ 
        cD211.054.96 & D96   & $96^3 \times 192$   & 1.900   & 0.05692(12) & 140.8(2)& 494      \\
        cE211.044.112 & E112   & $112^3 \times 224$   & 1.960   & 0.04892(11) & 136.5(2) &258\footnotemark[1]       \\ \hline \hline
        \end{tabular}}
\end{table}
\footnotetext[1]{This ensemble is currently under production and in this proceeding we show preliminary results for a subset of the configurations planned.}

The twisted-mass fermion discretization provides automatic $\mathcal{O}(a)$ improvement~\cite{Frezzotti:2000nk,Frezzotti:2003ni}. A clover term~\cite{Sheikholeslami:1985ij} is included to reduce isospin-breaking effects. The parameters of the ensembles used in this study are listed in Table~\ref{tbl:Ensembles}. Lattice spacings and pion masses for the B64, C80 and D96 are taken from Ref.~\cite{ExtendedTwistedMass:2022jpw} determined within the meson sector. These values agree with those determined from the nucleon mass~\cite{ExtendedTwistedMass:2021gbo}. The lattice spacing and pion mass for the E112 ensemble are taken from Ref.~\cite{ExtendedTwistedMassCollaborationETMC:2024xdf}.

Nucleon charges are extracted by combining two- and three-point nucleon correlation functions. To ensure consistent errors, statistics for three-point functions are increased as the  source-sink time separation is increased to keep the error approximately constant. The spectral decompositions of the two- and three-point functions are:\begin{gather} \label{eq:Twop} 
    C(\Gamma_0,\vec{p};t_s) = \sum_{i}^{\infty} c_i(\vec{p}) e^{-E_i(\vec{p})t_s}, \\  
    C^\mu(\Gamma_k,\vec{q};t_s,t_{ins}) = \sum_{i,j}^{\infty} A^{i,j}_\mu (\Gamma_k,\vec{q}) e^{-E_i(\vec{0})(t_s-t_{ins}) - E_j(\vec{q})t_{ins}}, 
    \label{eq:Threep}
\end{gather}  
where $t_s$ is the source-sink time separation and $t_{\rm ins}$ the
time separation between current insertion and source and $\vec{p}$ and
$\vec{q}$ are the two-point function sink momentum and three-point
function momentum transfer respectively. The axial case, for example,
is given as the ratio of the axial current three- to two-point
function for $\vec{p}=\vec{q}=0$:
\begin{gather} \label{eq:Ratio}  
    R^A_\mu(t_s,t_{ins}) = \frac{C_\mu^{3pt}(\Gamma_k;t_s,t_{ins})}{C^{2pt}(t_s)}
\ \xrightarrow[t_{ins}\rightarrow \infty]{t_s - t_{ins}\rightarrow \infty} g_A.  
\end{gather}
In the large time separation limit, the ratio converges to the corresponding charge, e.g., $g_A$. We analyze excited state contributions by performing two-state fits, where we explore a wide parameter space and average results using the Akaike Information Criterion (AIC)~\cite{Jay:2020jkz,Neil:2022joj}. The calculated nucleon matrix elements have been renormalized nonperturbatively by employing the RI$'$/MOM scheme followed by perturbative conversion to the $\overline{\rm MS}$ scheme at the reference scale of 2 GeV (see Ref.~\cite{Alexandrou:2024ozj} for more details).

\begin{figure}[t]
    \centering
    \begin{minipage}{\textwidth}
        \centering
        \includegraphics[width=\textwidth]{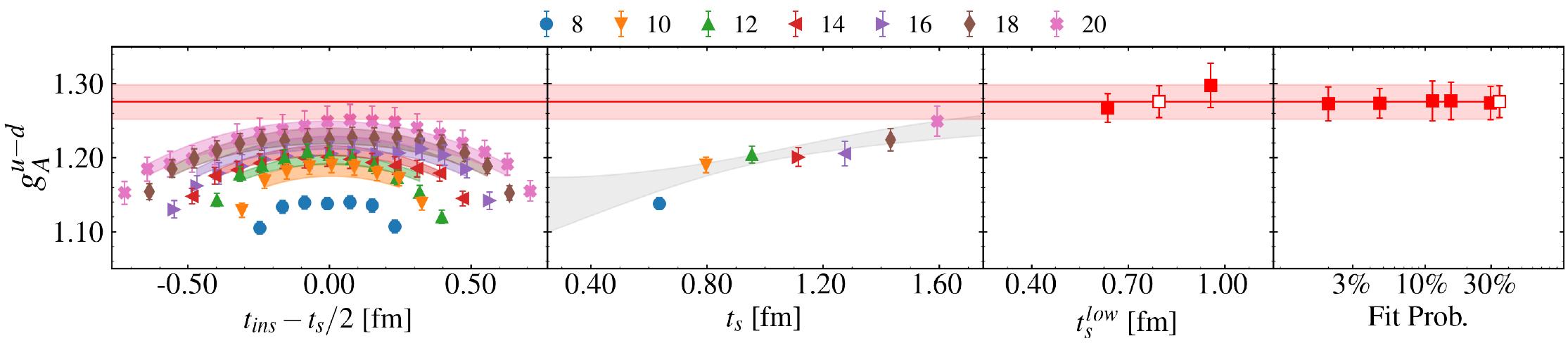}
        \vspace{0.5em}
        \includegraphics[width=\linewidth]{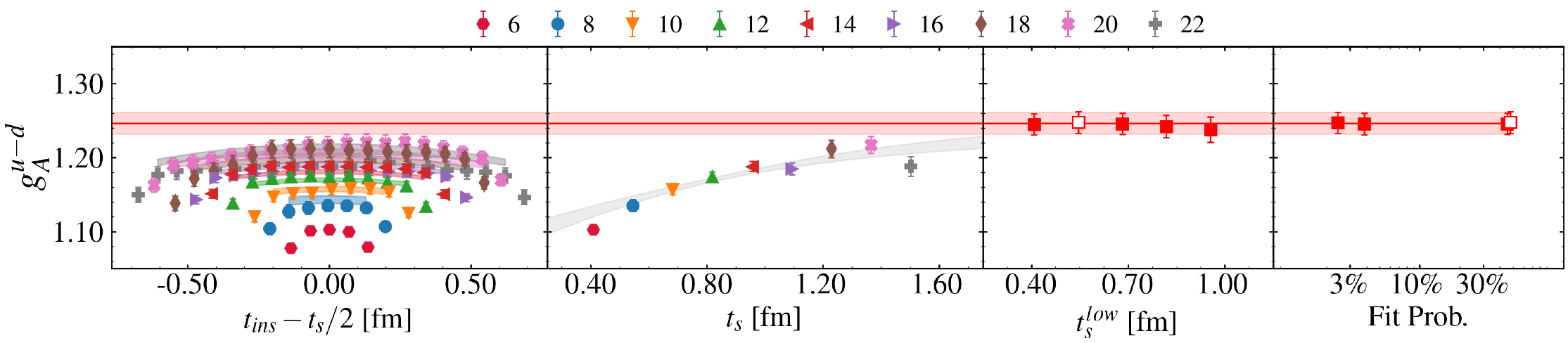}
        \vspace{0.5em}
        \includegraphics[width=\linewidth]{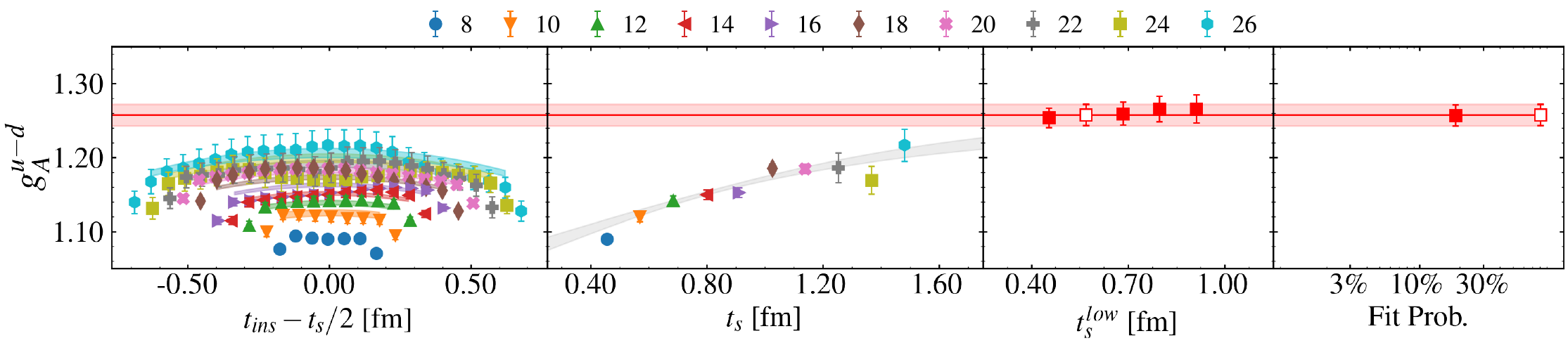}
        \caption{We present the ratio and fit results for all ensembles, for the isovector axial charge. In the legend, we give the symbols used
        to denote the various values of $t_s/a$. The top row shows the analysis for $\gaumd$ for the B64 ensemble, the middle row for C80 and the bottom for D96. The first column shows results on the ratio  versus $t_{ins}-t_s/2$. The horizontal bands are the model averaged  values. In the second column, we show the ratio versus $t_s$ for $t_{ins}=t_s/2$. The gray band is the result of the two-state fit model with the highest probability. In the third column, we show the extracted values as a function of the lowest value of $t_s$ included in the fits, where we give the results for the most probable model for a given value of $t_s^{\text{low}}$. In the last column, we present the weights for fit models whose probabilities exceed 1\%. The open symbols are the most probable models taking into account fits using  all $t_s^{\text{low}}$.}
        \label{fig:gA_umd}
    \end{minipage}
    \vspace*{-0.5cm}
\end{figure}

\section{Results}

In Fig.~\ref{fig:gA_umd} we present an example of a two-state fit analysis for the nucleon isovector axial charge $\gaumd$. The isovector axial-vector operator is given by:
\begin{gather}
    \ord^{u-d}_{A} \equiv  A_\mu = \bar{u} \gam \gfi u -\bar{d} \gam \gfi d,
\end{gather}
where $u$ and $d$ are the up and down quark fields, respectively.
For the axial case, due to chiral perturbation theory arguments~\cite{Bar:2018xyi,Bar:2016uoj}, to better estimate the isovector axial charge,  we include in the fit the three-point function of the temporal component of the axial-vector current with one unit of momentum transfer,
$C^0(\Gamma_k, \frac{2\pi}{L}\hat{k}; t_s, t_{ins})$ together with the corresponding two-point function, to help extract the excited state energies $E_i(\vec{q})$, where $L$ is the
spatial length of the lattice and $\hat{k}$ a unit vector in the $k$
spatial direction. 

In our fitting procedure, we vary the smallest value of the sink time
$t_s$ we use for the fitting of the ratio, $t_s^{\text{low}}$, as well
as the lowest time slice in the fitting of the two-point functions,
$t_{2pt}^{low}$. We also vary the number of insertion time slices that
we keep in the fits. We use $t_{ins} \in
[t_{ins,0},t_s-t_{ins,s}]$. For the charges we work at $Q^2=0$ so we
fix $t_{ins,0}=t_{ins,s}$, removing an equal number of time slices on
the source and the sink sides. The extracted values
show almost no dependence on on $t_s^{\text{low}}$. The results in the
third column of Fig.~\ref{fig:gA_umd} represent the highest
probability model for each $t_s^{\text{low}}$, varying
$t_{\text{ins,0}}$ and $t_{2\text{pt}}^{\text{low}}$.

\begin{figure}[htbp]
        \vspace*{-0.3cm}
	\includegraphics[width=\linewidth]{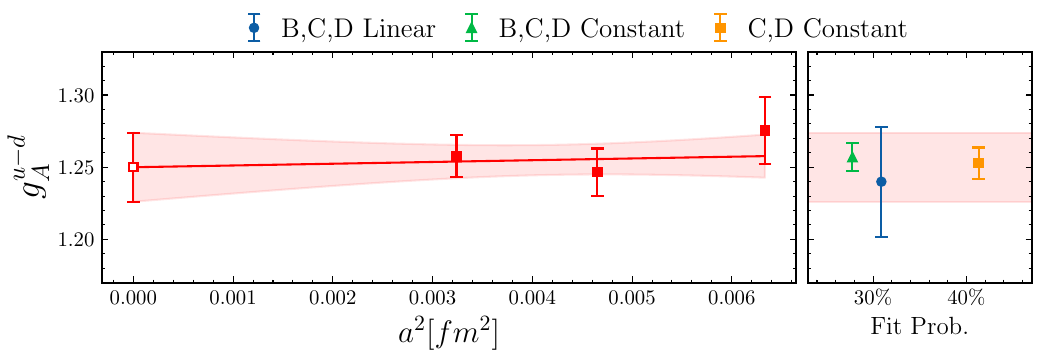}\\
	\vspace*{-0.5cm}
	\caption{In the left panel, we show the continuum limit of the nucleon isovector axial 
          charge (open symbol and band) extrapolated using the B64, C80 and D96 ensembles 
          (filled symbols). The extrapolation is the result of a model
          average, which combines linear and constant fits as explained
          in the text. In the right panel, we show the weights for each type of fit, namely the linear extrapolation is represented by a blue circle, the constant extrapolation including all three ensembles by a green triangle and the constant extrapolation using the C90 and D96 ensembles  by an orange square. The red band is the model average value from the three fits.}
	\vspace*{-0.4cm}
        \label{fig:gA_extrap}
\end{figure}

To extrapolate to the continuum limit ($a \to 0$), we use the results from the ensembles B64, C80 and D96. We carry out three types of extrapolation and evaluate a combined systematic and statistical error via a model average over the three fits. Namely, we use a linear fit in $a^2$ and a constant fit either using all three ensembles or when omitting the coarser ensemble, B64. A strong dependence on the lattice spacing will result in a model average favoring the linear fit, while a mild $a^2$ dependence will lead to a model average favoring the two constant fits. The continuum limit extrapolation for the isovector axial charge is shown
in Fig.~\ref{fig:gA_extrap}, along with the weights for each type of extrapolation. In Table~\ref{tbl:Res} we present our continuum limit results for the axial, scalar and tensor 2-, 3- and 4- flavor isovector and isoscalar combinations, as well as the single flavor charges for each case. All results per ensemble and after continuum extrapolation can be found in Ref.~\cite{Alexandrou:2024ozj}.

\begin{table}[htbp]
	\centering
	{\small
		\renewcommand{\arraystretch}{1.0}
		\renewcommand{\tabcolsep}{7.0pt} 
        \caption{Values for the 2-, 3-, and 4-flavor isovector and isoscalar combinations (top) and the extracted single flavor charges (bottom) in the continuum limit, using the model average strategy described
        in the text.}
        \label{tbl:Res}
        \vspace*{-0.2cm}
        \begin{tabular}{c|c|c|c|c|c}
        \hline \hline
         & $u-d$ & $u+d$ & $u+d-2s$ & $u+d+s-3c$ & $u+d+s+c$ \\ \hline
        \multicolumn{1}{c|}{$g_A$} & 1.250(24) & 0.423(33) & 0.490(20) & 0.343(55) & 0.382(70) \\ \hline
        \multicolumn{1}{c|}{$g_S$} & 1.08(31) & 11.5(2.2) & 11.2(2.1) & 11.4(2.1) & 12.2(2.2) \\ \hline
        \multicolumn{1}{c|}{$g_T$} & 0.955(29) & 0.561(34) & 0.561(33) & 0.569(37) & 0.557(34) \\ \hline \hline
        \end{tabular}
        \\[0.1cm]
      \begin{tabular}{c|c|c|c|c}
        \hline \hline
         & $u$ & $d$ & $s$ & $c$ \\ \hline
        \multicolumn{1}{c|}{$g_A$} & 0.832(28) & -0.417(22) & -0.037(18) & 0.003(13)  \\ \hline
        \multicolumn{1}{c|}{$g_S$} & 6.4(1.1) & 5.30(98) & 0.16(37) & 0.09(26) \\ \hline
        \multicolumn{1}{c|}{$g_T$} & 0.756(29) & -0.196(12)  & -0.0009(11) & -0.0028(26)\\ \hline \hline
        \end{tabular}
	}
\end{table}

In this work, we also calculated the nucleon $\sigma$-terms, which are defined as 
\begin{gather}
    \sigma^f= m_f \braket{N|\bar{\psi}_f  \psi_f|N}~,~~\sigma^{u+d}=m_{ud} \braket{N|\bar{u}u+\bar{d}d|N},
\end{gather}
where $m_f$ is the quark mass for a given flavor $f$, $m_{ud}$ is the average light quark mass, and $\ket{N}$ is the nucleon state. The value of $\sigma^{u+d}$, also referred to as $\sigma^{\pi N}$, is determined from phenomenological analyses using experimental inputs. These quantities are fundamental in QCD, providing insights into the quark content of the nucleon and serving as a measure of chiral symmetry breaking.

The nucleon $\sigma$-terms are extracted from the scalar matrix elements, including disconnected quark loops. The twisted mass formulation simplifies renormalization compared to standard Wilson fermions, as it eliminates additive mass renormalization and ensures that the multiplicative renormalization of the scalar current and quark mass cancel. Our results for $\sigma^{\pi N}$, $\sigma^s$, and $\sigma^c$ are presented in Table~\ref{tbl:sigma_terms}.

\begin{table}[htbp]
	\centering
	{\small
		\renewcommand{\arraystretch}{1.3}
		\renewcommand{\tabcolsep}{5.0pt}
        	\caption{Results for the  nucleon $\sigma$-terms (in MeV) in the continuum limit. For $\sigma_{\pi N}$ and  for $\sigma_s$ we follow the same extrapolation procedure as described in Fig.~\ref{fig:gA_extrap}, while for $\sigma_c$ we use a single constant extrapolation.}
	\label{tbl:sigma_terms}
		\begin{tabular}{c|c||c|c||c|c}
        \hline \hline
               $\sigma^{\pi N}$ & 41.9(8.1) & $\sigma^{s}$ & 30(17) & $\sigma^{c}$& 82(29)       \\ \hline \hline
        \end{tabular}
	}
	\vspace*{-0.4cm}
\end{table}

\begin{figure}[htb]
    \centering
    \begin{minipage}{\textwidth}
        \centering
        \includegraphics[width=\textwidth]{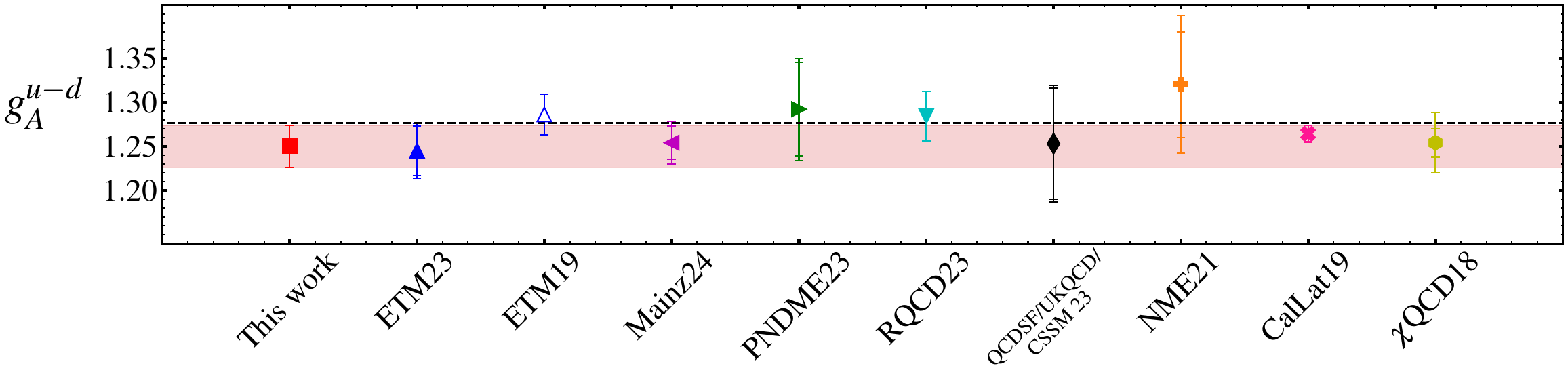}
        \vspace{0.5em}
        \includegraphics[width=\linewidth]{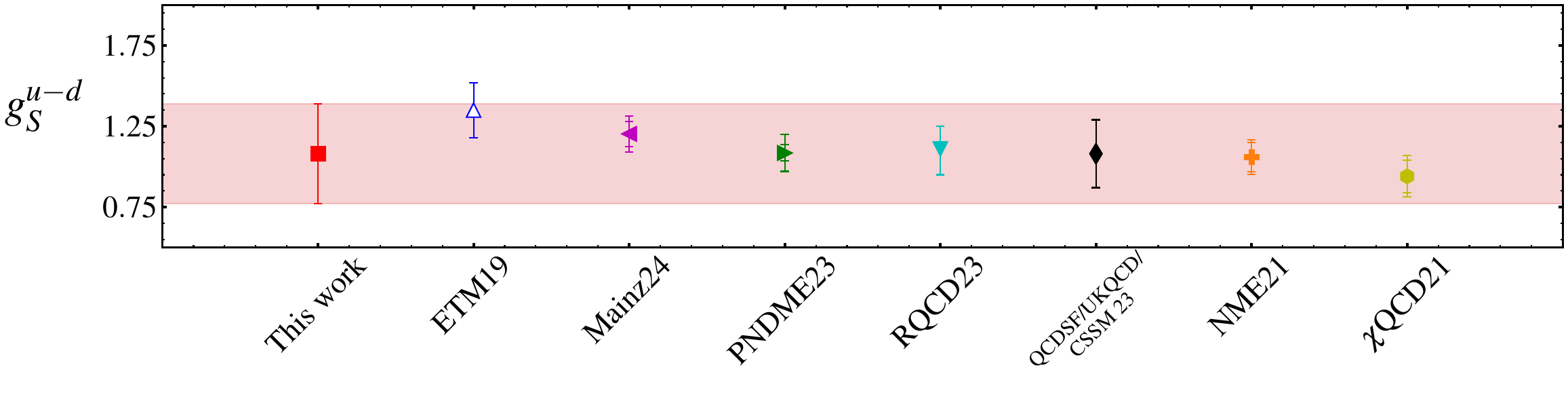}
        \vspace{0.5em}
        \includegraphics[width=\linewidth]{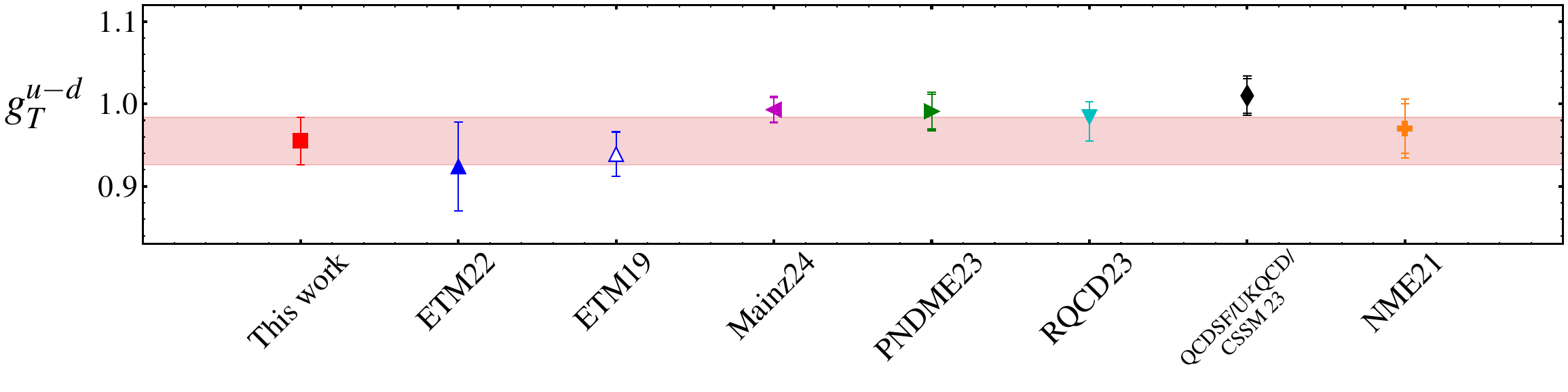}
        \caption{Comparison of the results of this work with other lattice QCD results, for the isovector axial, scalar and tensor charges. Our results are shown with the red square and red error band.  The blue triangles show previous ETMC results, for $\gaumd$~\cite{Alexandrou:2023qbg} and for $g_T^{u-d}$~\cite{Alexandrou:2022dtc}, while Ref.~\cite{Alexandrou:2019brg} gives results  on all three isovector charges including $g_s^{u-d}$ for the B64 ensemble. Open symbols represent results without a continuum limit extrapolation. The magenta triangles show recent results from the Mainz group~\cite{Djukanovic:2024krw}, the green triangles from PNDME~\cite{Jang:2023zts}, the cyan triangles from RQCD~\cite{Bali:2023sdi}, the black diamonds from the QCDSF/UKQCD/CSSM collaboration~\cite{QCDSFUKQCDCSSM:2023qlx}, the orange crosses from NME~\cite{Park:2021ypf}, the pink cross from CalLat~\cite{Walker-Loud:2019cif} and the yellow hexagons from $\chi$QCD~\cite{Liang:2018pis,Liu:2021irg}. For $\gaumd$, the dashed line represents the experimental value~\cite{Markisch:2018ndu}.}
        \label{fig:umd_comb}
    \end{minipage}
    \vspace*{-0.3cm}
\end{figure}

In Fig. \ref{fig:umd_comb}, we compare  our results for the isovector charges with  recent results from other lattice QCD studies, as well as with previous analyses of
these quantities by ETMC. The results presented in this work are the only ones obtained by taking the continuum limit using ensembles simulated directly at the physical pion mass. In contrast, all other  collaborations use ensembles with pion masses heavier than physical or combine one or two physical point ensembles with heavier-than-physical ones. Since results at the physical point typically have larger statistical uncertainties, their extrapolations may be more influenced by the heavier-than-physical ensemble data.

We observe very good agreement among lattice QCD results by different collaborations for all isovector charges. Moreover, our value for the isovector axial charge $\gaumd$, is compatible with the experimental value~\cite{Markisch:2018ndu}.

\clearpage
\subsection{Preliminary results using the E112 ensemble with $a\approx 0.05$ fm.}

In this section, we present preliminary results for our new physical point ensemble, simulated at a finer lattice spacing ($a\approx 0.05$ fm), that will aid in the continuum limit extrapolation of all quantities. The details of the ensemble are listed in Table~\ref{tbl:Ensembles}.

\begin{figure}[ht!]
	\includegraphics[width=\linewidth]{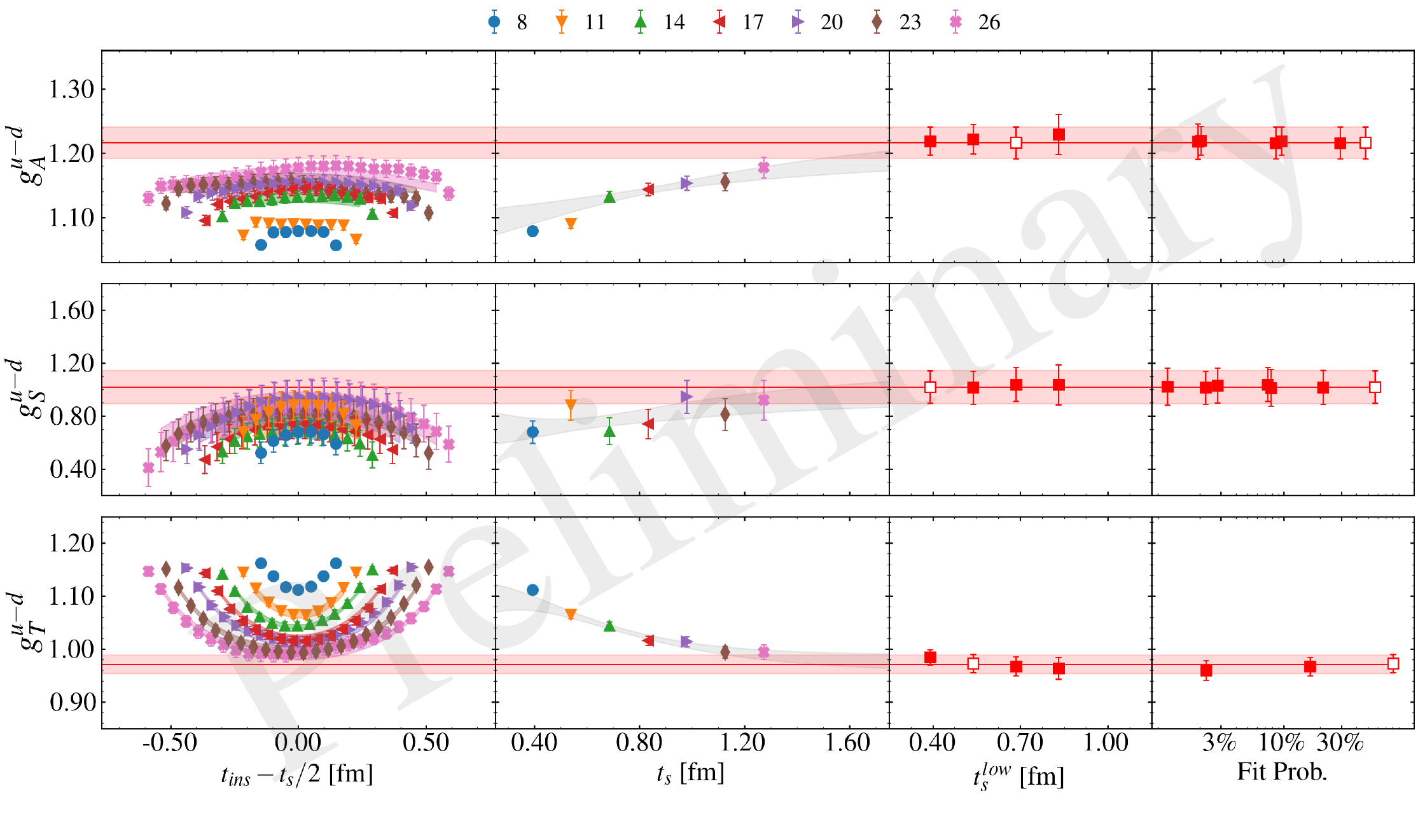}\\
	\vspace*{-0.3cm}
	\caption{Ratio and preliminary fit results for the currently available data of the E112 ensemble, for the isovector charges. The notation is the same as in Fig. \ref{fig:gA_umd}.}
	\vspace*{-0.4cm}
        \label{fig:umd_E112}
\end{figure}

In Fig.~\ref{fig:umd_E112}, we show the ratios and preliminary fits for the isovector axial, scalar and tensor charges for the E112 ensemble. We observe a good signal to noise ratio for the analyzed configurations. We aim to use more than 500 configurations, increase statistics and the value of $t_s$. Moreover, we will calculate the disconnected contributions and perform the continuum limit extrapolation using all four ensembles. 

\section{Conclusions}
We present results on the nucleon axial, scalar and tensor charges, as well as on the nucleon $\sigma$-terms, using three $N_f=2+1+1$  twisted mass clover-improved fermions ensembles, with quark masses tuned to reproduce their physical values. This enables us, for the first time, to determine these charges and $\sigma$-terms  at the continuum limit  using only physical point ensembles, avoiding any chiral extrapolations. Our future goal is to include the fourth ensemble with a finer lattice spacing of  
$a\approx 0.05$ fm for which we include preliminary results in this proceeding. This finer ensemble will further refine our continuum  limit extrapolations and increase the accuracy of our final values.

\acknowledgments
We thank the ETM collaboration for their support and particularly acknowledge Matteo Di Carlo, Antonio Evangelista, Roberto Frezzotti, Giuseppe Gagliardi, and Vittorio Lubicz for discussions and crosschecks on renormalization factors. We are also grateful to Georg Bergner, Petros Dimopoulos, Bartosz Kostrzewa, and Urs Wenger for their contributions to the $N_f=4$ gauge ensembles.  

C.A. and G.K. acknowledge partial support from the European Joint Doctorate AQTIVATE (Grant Agreement No.~101072344). Y.L. and Ch.I. are supported by the Excellence Hub project ``3D-nucleon'' (id EXCELLENCE/0421/0043) co-financed by the European Regional Development Fund and the Republic of Cyprus through the Research and Innovation Foundation and by the University of Cyprus projects "Nucleon-GPDs" and "PDFs-LQCD". S.B. and G.K. acknowledge support from the Excellence Hub project ``NiceQuarks'' (id EXCELLENCE/0421/0195). S.B. and J.F. are supported by the Inno4scale project, which received funding from the European High-Performance Computing Joint Undertaking (JU) under Grant Agreement No.~101118139. J.F. also acknowledges support from the DFG research unit FOR5269 ``Future methods for studying confined gluons in QCD'' and the Next Generation Triggers project (\url{https://nextgentriggers.web.cern.ch}).

We thank the Gauss Centre for Supercomputing e.V.~(\url{www.gauss-centre.eu}) for access to SuperMUC-NG at the Leibniz Supercomputing Centre, JUWELS~\cite{JUWELS} and JUWELS Booster~\cite{JUWELS-BOOSTER} at the Jülich Supercomputing Centre (JSC). Results were partially created within the JUWELS Booster EA program with assistance from the JUWELS Booster Project Team (JSC, Atos, ParTec, NVIDIA). We acknowledge the Swiss National Supercomputing Centre (CSCS) and the EuroHPC Joint Undertaking for awarding access to the LUMI supercomputer, owned by the EuroHPC Joint Undertaking, hosted by CSC (Finland) and the LUMI consortium, through the Chronos programme under project IDs CH16-CYP. We are grateful to CINECA and the EuroHPC JU for awarding access to supercomputing facilities hosted at CINECA and to Leonardo-Booster through the Extreme Scale Access Call grant EHPC-EXT-2024E01-027. Finally, we acknowledge computing time on Cyclone at the Cyprus Institute (project IDs P061, P146, and pro22a10951).

\newpage

\bibliographystyle{JHEP}
\bibliography{refs}

\end{document}